\documentclass[
reprint,
superscriptaddress,
amsmath,amssymb,
aps,
fleqn,
prb
]{revtex4-1}

\usepackage{graphicx}
\usepackage{dcolumn}
\usepackage{bm}

\begin{document}


\title{Local structure and defect segregation on the tilt grain boundaries in silicon}

\author{Vitaly Yu. Lazebnykh}
\affiliation{ Irkutsk State Technical University, 83 Lermontov Street, 664074 Irkutsk, Russia}

\author{Andrey S. Mysovsky}
\email{mysovsky@gmail.com}
\affiliation{ Irkutsk State Technical University, 83 Lermontov Street, 664074 Irkutsk, Russia}
\affiliation{A.P. Vinogradov Institute of Geochemistry SB RAS, 1a Favorsky Street, 664033 Irkutsk, Russia}

\date{\today}

\begin{abstract}
We present the results of atomistic and {\it ab initio} simulation of several different tilt grain boundaries (GB) in silicon. The boundary structures obtained with genetic algorithm turned out to have no coordination defects, i.e. all silicon atoms restored their tetrahedral coordination during the structure optimisation. That concerns previously known symmetric $\Sigma 5~(130)$, $\Sigma 3~(211)$ and $\Sigma 29~(520)$ boundaries as well as previously unknown asymmetric $\Sigma 9~(\overline{2}55)/(\overline{2}11)$, $\Sigma 3~(\overline{2}55)/(211)$ and $\Sigma 13~(790)/(3~11~0)$  structures. We have performed an extensive study of defect segregation on the boundaries, including neutral vacancy and carbon, phosphorus and boron impurities. A clear correlation between the segregation energy of the defect and local geometry of the boundary site where the defect is segregated has been revealed. We suggest a simple purely geometric model for evaluation of approximate segregation energies of the listed defects.
\end{abstract}


\maketitle


\section{\label{sec:intro}Introduction}

Grain boundaries (GBs) drastically affect the electronic properties of polycrystalline silicon used for solar cell and electronic devices. The boundary related gap states greatly contribute to the recombination of charge carriers~\cite{fraas}, thus leading to the decrease  of the minority carriers lifetime and diffusion length and the deterioration of device performance. A charged boundary, e.g. the one acting as a sink for charged defects or having enough traps for charge carriers, can create a depleted layer and become itself an electrostatic  barrier for charge carriers~\cite{seto-barriers}. 

One possible source of gap states at GBs is the presence of coordination defects. By coordination defects we shall further mean any deviation from fourfold (in other words, tetrahedral) coordination of silicon atoms at the boundary. Such defects might exist even in pure boundaries containing no impurities and having favourable energies and stable structures. Despite the extensive experimental and computational studies the presence of coordination defect at different boundary types remains controversial.
For example, the authors of recent work (Ref.~\onlinecite{passivate-limit}) take it as something obvious that the geometries of grain boundaries are highly strained, resulting in deep levels in the band gap.

This is not always the case at least for twin and symmetric tilt boundaries. Yet in the very first theortical work on the structure of GBs in silicon by M\"oller~\cite{moller}, despite very simple empirical approach to the GB energy calculation and absence of automated geometry optimisation, several perfectly bonded structures of $\langle 011 \rangle$ symmetric tilt GBs have been revealed. After the {\it Structural Unit Model} has been proposed in the seminal paper of Sutton and Vitek~\cite{sutton-sum1,*sutton-sum2} it became possible to find the perfectly bonded low energy models of symmetric tilt boundaries even in early computational studies~\cite{kohyama-tilt1,*kohyama-tilt2,sutton130-1,*sutton130-2,nikolaeva-S5-210}.
In later simulations (for example, Refs.~\onlinecite{huang-tilt130,morris-sigma13}) many more of such models have been established. One can find an extensive review of computational studies in Ref.~\onlinecite{kohyama-review}. In diamond which has the same crystalline structure as silicon the situation is similar despite the higher stiffness of C-C compared to Si-Si bonds.
The family of symmetric $\langle 001 \rangle$ tilt GBs in diamond~\cite{shenderova-diamond-tilt} has been found to have no coordination defects with few exceptions.
 Tight binding (TB) study~\cite{morris-tb-diamond} of symmetric $\langle 011 \rangle$ tilt boundaries revealed the bond reconstruction between threefold-coordinated atoms leading, again, to perfectly bonded structures.

For twist grain boundaries the situation is different. Most of proposed models for silicon~\cite{wolf-twist,twist87,twist89,twist90,cleri-twist,cleri-twist2}, diamond~\cite{keblinski98} and ultrananocrystalline diamond~\cite{zapol-tbmd-diamond} contain coordination defects. Besides coordination defects, partial amorphisation of intergranular layer is often attributed to these boundaries~\cite{cleri-twist}.

However, van Alfthan {\it et al}~\cite{alfthan2006-twist} has pointed out that the structure search algorithms used in previous studies did not sample the configurational space correctly. The key point of a newly suggested protocol  was systematic removal of atoms from the boundary.
With this new protocol they were able to find ordered structures without coordination defects even for twist boundaries.
Steneteg {\it el al}~\cite{steneteg-diamond} later underlined the importance of this protocol in the study of $\Sigma 5~(001)$ twist GB in diamond, however, their lowest energy models contained coordination defects.

Simplifying the situation, one may conclude that there are two positions. According to the first one, pure boundaries in silicon in their stable configurations are ordered and nearly perfectly bonded. According to the second one, amorphisation and bonding defects are inherent for GBs in silicon. The right answer might include both those positions. For example, Kutsukake {\it et al}~\cite{struct-imperfection} artificially synthesised $\Sigma 5~(310)$ boundaries with small angular deviation from ideal misorientation angle $\alpha=36.87^\circ$. For zero angular deviation the boundary had shown no electrical activity at all, while for larger angular deviations it was growing. In other words, it is possible that there exist ``ideal'' boundaries, ordered, perfectly bonded and electrically inactive, while small angular deviation from them create amorphisation, coordination defects and gap states.


The second possible source of gap states at the boundary is the segregation of point defects. It is, in fact, very difficult to distinguish experimentally between impurity and coordination defects related gap states.
This circumstance becomes additional obstacle for clarifying the structure of GBs and their segregation and electronic properties.

The behaviour of impurities on the boundary is by no means less controversial then the structure of boundaries themselfs. For example, electrical activity of GBs, initially weak in undoped samples, turned out to become significant upon iron doping in electron beam induced current measurements~\cite{chen-ebic}. Moreover, the dependence of electrical activity and, presumably, iron segregation on the boundary type has been observed. 
On the contrary, the results of microwave detected photoconductivity and  deep level transient spectroscopy~\cite{pss-recombination} as well as photoluminescence and minority carriers lifetime measurement~\cite{sugimoto-pl} suggest that  the dominant recombination path is via decorated dislocation clusters within grains with little contribution to the overall recombination from GBs.

It should be noted that gap states on the grain boundaries is not the only possible thing leading to their electrical activity. Pure electrostatic effects, for example, for the boundaries segregating charged defects, might be significant as well and lead to electrical activity. Photothermal signal indicating the presence of carriers recombination has been found to correlate with electrostatic potential on GBs\cite{hara-phototerm}. The potential barrier energy of 0.22~eV determined~\cite{hashimoto-carrier-mobility} for twin $\Sigma 3~(111)$ boundary from measurements of carrier mobility temperature dependence provides characteristic for the order of magnitude of the boundary potential.

A number of computational studies deal with impurities segregated at grain boundaries. Iron at $\Sigma 5~(310)$ GB in silicon~\cite{fe-sigma5}, Ga and As at the same boundary in Ge~\cite{arias-joannopoulos}, neutral vacancy at $\Sigma 3~(211)$ GB~\cite{vac-3-112} -- all this works predict configurations, segregation energies, preferable segregation sites and electronic structure of impurities. However, all these models apply only to one specific boundary they were established for and provide little general information about segregation of these impurities, i.e. information valid for a variety of GBs.

This is not just a question of structure complexity, it is also a question of language. What language do we have to describe defects in irregular structures? Sometimes it is not even clear what should and what should not be considered as ``defect'' in a structure which is itself inherently defective from the crystalline point of view (an example will be given in sec.~\ref{sec:vacancy}). The crystalline concept of point defect having the same configuration and same properties in any of lattice sites does not work well any more and should be supplemented, if not replaced, with something else.

The most obvious way to deal with this difficulty is to introduce local structural parameters characterising each site of irregular structure. Than one can search for correlations between the properties of the defect of concern and local structure of the sites where it is placed. In this work we suggest a parameter set exhibiting clear correlation with the energies of defects segregated at tilt grain boundaries in silicon.

In present paper we report atomistic and {\it ab initio} simulations results for  several symmetric and asymmetric tilt grain boundaries in silicon. Having established their structures we have considered segregation of such defects as carbon, phosphorus and boron impurities and neutral vacancy in different sites of the boundaries. Segregation energies and electronic structure of the defects is discussed.


\section{\label{sec:calc}Calculation details}

To search for low energy GB structures we have implemented genetic algorithm (GA)~\cite{Jensen2006-342} in Python scripting language. In genetic algorithm a pool of candidate structures is initially created from random structures. Then at each step of algorithm the most successful structures, i.e. those possessing lowest energies, give ``offspring'' by mixing their parts in a new structure. Certain amount of ``offspring'' and ``parent'' structures form new structure pool. Genetic algorithm might also include random mutations at any step. Each step of GA includes also conventional geometry optimisation usually done with conjugate gradients.

In our implementation of GA the pool contained 100 candidate structures. The structure optimisation was limited to 10~{\AA} thick layer containing the grain boundary while the grains outside this layer remained intact, although the rigid body translation of one grain relative to another was allowed. The ``offspring'' was produced as follows: two parent structures were cut but by randomly placed plane perpendicular to the boundary and the parts of these structures were combined.  

For the purpose of accurate evaluation of GB energies the {\it double boundary} model has been utilised in both atomistic and {\it ab initio} calculations. In double boundary model the supercell contains two identical boundaries related to each other with inversion symmetry operation. Unlike a single boundary in a slab, the double boundary supercell does not have interfaces with vacuum possessing unknown energy and bringing uncontrollable error to the GB energy.

The energy evaluation and geometry optimisation of grain boundary structures has been performed at atomistic level with Tersoff~III many-body potentials~\cite{tersoff1}. The same level of theory has been applied to model carbon impurity segregation~\cite{tersoff2}. Tersoff potentials are known to provide reliable GB geometries and in vast majority of cases preserve correct energy ranking order, however, GB energies calculated with Tersoff potentials are systematically overestimated by approximately 30\% compared to {\it ab initio} results~\cite{keblinski98}. General utility lattice program (GULP 3.4)~\cite{gulp} has been used for atomistic calculations. 

 For phosphorus and boron as well as for the neutral vacancy we have performed {\it ab initio} calculations at GGA level of density functional theory. We used PBE functional~\cite{pbe} in combination with PAW pseudopotentials~\cite{paw} implemented in  Vienna {\it ab initio} simulation package (VASP 5.2)~\cite{vasp}. The energy cutoff for plane wave basis set of 245.3~eV has been applied to the vacancy and pure boundary calculations, while for P and B the cutoff values were 270.0 and 318.6~eV, respectively. These cutoff values have been chosen because they are minimal recommended for corresponding PAW pseudopotentials. Monkhorst-Pack~\cite{monkhorst-pack} k-points mesh has been constructed for Brillouin zone sampling. In calculations involving evaluation of gradients (e.g. geometry optimisation) the mesh dimension was $3\times 3 \times 3$, while for density of states (DOS) calculations it was $5\times 5 \times 5$.


\section{\label{sec:struct}Grain boundary structures}

We have calculated symmetric $\Sigma 5 (130)$, $\Sigma 3 (211)$ and $\Sigma 29 (520)$ and  asymmetric $\Sigma 9(\overline{2}55)/(\overline{2}11)$, $\Sigma 3(\overline{2}55)/(211)$  and $\Sigma 13 (790)/(3~11~0)$ grain boundaries. Some of these results hav been already reported~\cite{self1} in more details. The listed symmetric structures are already well known~\cite{kohyama130,kohyama211} and allow us to validate  our implementation of the genetic algorithm. These particular asymmetric structures, on the other hand, has been chosen because they require not very large supercells for their modelling and their calculation is feasible. All grain boundary energies are summarised in Table~\ref{tbl:GB_energy}.

\begin{table}
\caption{Energies of calculated grain boundaries}
\label{tbl:GB_energy}
\begin{ruledtabular}\begin{tabular} {c d c l c d}
 axes  &  \multicolumn{1}{c}{$\theta^{\circ}$} & $\Sigma$ & \multicolumn{1}{c}{plane(s)}  &  
 type$^*$ & \multicolumn{1}{c}{$E_{GB},J/m^2$} \\
\hline
$\langle 001 \rangle$ & 22.61 & 13 & (790)/(3 11 0)                        &  a  & 1.12 \\
                      & 36.89 & 5  & (130), model A                          &  s  & 0.68 \\
                      & 36.89 & 5  & (130), model B                          &  s  & 0.63 \\
                      & 43.6  & 29 & (520)                                   &  s  & 0.77 \\
                      & 43.6  & 29 & (520), metastbl.                        &  s  & 1.27 \\
$\langle 011 \rangle$ & 38.94 & 9  & ($\overline{2}$55)/($\overline{2}$11) &  a  & 1.0 \\
                      & 70.52 & 3  & ($\overline{2}$55)/(211)              &  a  & 0.37 \\
                      & 70.52 & 3  & (211)                                   &  s  & 0.69 \\
\end{tabular}\end{ruledtabular}
$^*$ s - symmetric, a - asymmetric GB
\end{table}

For the $\Sigma 5~(130)$ several subsequent runs of GA were able to recover both ``model A'' and ``model B'' described by Kohyama {\it et al}~\cite{kohyama130}. The model B has zigzag structure and lower energy than model A with a mirror-plane symmetry.  The GB energies of 0.68 and 0.63~$J/m^2$ obtained with Tersoff potentials are to be compared with 0.42 and 0.29~$J/m^2$ obtained in cited work. The energy ranking order is the same, the qualitative agreement can be admitted, and the quantitative accuracy was not even expected.

For $\Sigma 29~(520)$ boundary the first run of GA ended up with amorphous-like configuration with relatively high energy of 1.27~$J/m^2$. Although perfectly bonded, this configuration is not energetically favourable and is marked as ``metastable'' in Table~\ref{tbl:GB_energy}. Another GA run has found the mirror-plane lowest energy structure. This difficulty demonstrates that even with sophisticated algorithms like GA the problem of searching for complicated irregular structures is far from being fully resolved.


\begin{figure}
\center
\begin{tabular}{cc}
\includegraphics[width=0.5\columnwidth]{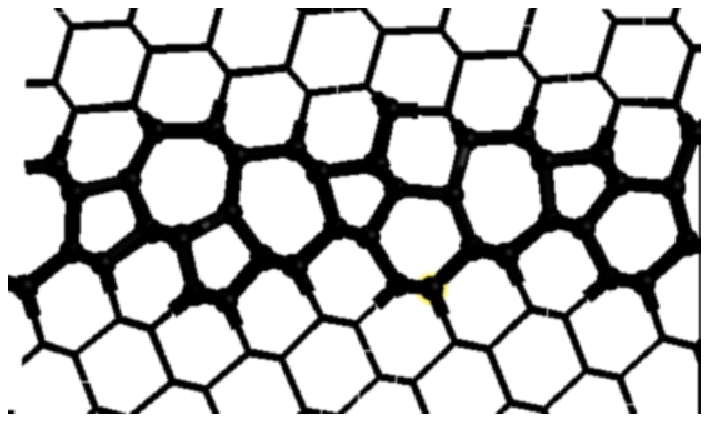} &
\includegraphics[width=0.45\columnwidth]{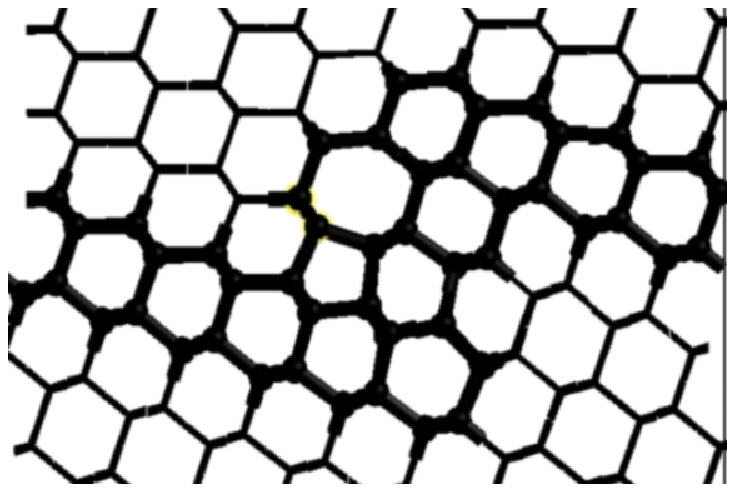}\\
a) & b) \\
\multicolumn{2}{c}{\includegraphics[width=0.9\columnwidth]{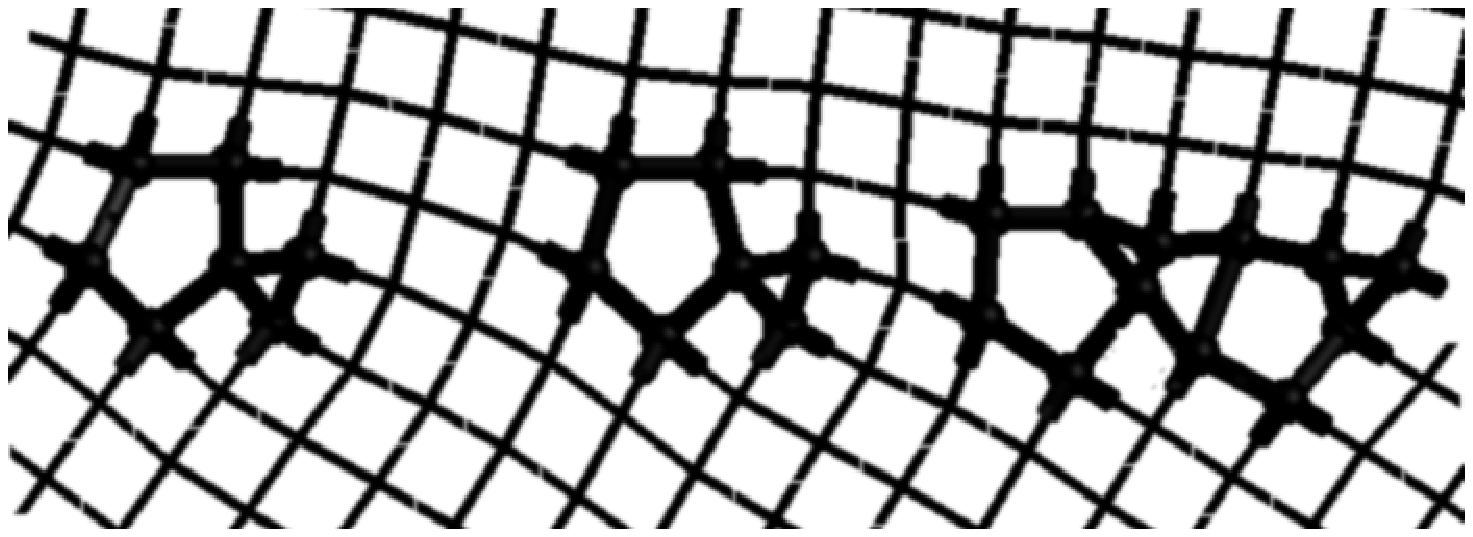}} \\
\multicolumn{2}{c}{ c) } 
\end{tabular}
\caption{\label{fig:structures} Lowest energy structures of asymmetric grain boundaries a)  $\Sigma 9  (\overline{2}55)/(\overline{2}11)\langle 011\rangle$; b) $\Sigma 3(\overline{2}55)/(211)\langle 011\rangle$; c) $\Sigma 13  (790)/(3~11~0) \langle 001\rangle$ }
\end{figure}

Asymmetric tilt boundary structures obtained in this work are shown on fig.~\ref{fig:structures}. The $\Sigma 13~(790)/(3~11~0)\langle 001\rangle$ and $\Sigma 9~(\overline{2}55)/(\overline{2}11)\langle 011\rangle$ GBs have rather high energies of 1.12 and 1.0~$J/m^2$. They are perfectly bonded and consist of the same structural elements as symmetric boundaries with corresponding rotation axes ($\langle 001 \rangle$ and $\langle 011 \rangle$), however, taking into account the behaviour of GA described above, it cannot be guarantied that those structures are indeed lowest energy ones. For the boundary $\Sigma 3~(\overline{2}55)/(211)\langle 011\rangle$ we have found surprisingly low energy of 0.37~$J/m^2$.  As Tersoff potentials usually overestimate GB energies, the exact value can be lower by factor of 1.5-2. Another interesting feature of this GB is its zigzag structure (fig. 1b).

\section{\label{sec:segreg}Defect segregation}

\subsection{\label{sec:vacancy} The neutral vacancy}

Although being one of the simplest intrinsic defects in silicon, the neutral vacancy puts considerable challenge for computational studies. First of all, atomistic simulation should be excluded due to its failure to reproduce Jahn-Teller distortion of the vacancy coordination lowering the symmetry from $T_d$ to $D_{2d}$. Hereafter by ``vacancy coordination'' we mean four nearest neighbours of the removed atom. Second, {\it ab initio} calculations of the vacancy require rather large supercell (not less than 128 atoms according to Puska {\it et al}~\cite{puska-vacancy}). In smaller supercells vacancy interaction with its images in neighbouring supercells through long range lattice deformation leads to significant error in both geometries and formation energies.

The difficulties in theoretical description of vacancy are to be seen in the example of formation energy. The values calculated by different authors range from 2.86 to 4.12~eV (see the comparison in Ref.~\onlinecite{fedwa-vacancy}) depending on the supercell size, Brillouin zone sampling and DFT flavour used. It worth to note that all those calculations were performed in very similar framework of plane wave LDA or GGA density functionals. In other words, accurate calculation of the formation energy is problematic and only semiquantitative level of description is available. The experimental value determined from positron-lifetime measurements~\cite{vacancy-positron-lifetime} is 3.6~eV.

Our calculations provide similar values for the formation energy of the bulk vacancy. In supercell of 250 atoms and $3\times 3\times 3$ k-mesh it was 3.3~eV. 
The tetrahedral coordination of the vacancy undergoes an asymmetric deformation. Silicon atoms in two opposite pairs approach each other to the distance about 2.9~$\AA$ while the distance between pairs remains 3.35~$\AA$. Again, our results comply with the picture of previous works~\cite{puska-vacancy,fedwa-vacancy}.

Next, we have calculated the {\it segregation energy} for the vacancy in all possible boundary sites for $\Sigma 5~(130)\langle 001\rangle$, $\Sigma 29~(520)\langle 001\rangle$  and $\Sigma 3~(211)\langle 011\rangle$ GBs. The supercells for these GBs contained 248, 228 and 400 atoms, respectively. 

\begin{figure}
\center
\includegraphics[width=\columnwidth]{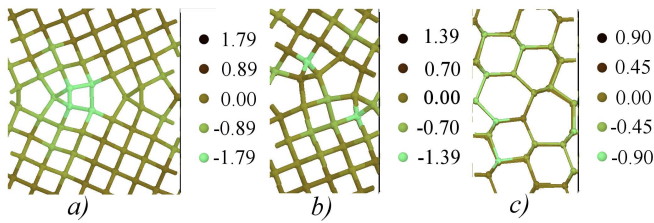}
\caption{\label{fig:smapV5} Segregation energy maps of the vacancy on $a)$~$\Sigma 29~(520)\langle 001\rangle$, $b)$~$\Sigma 5~(130)\langle 001\rangle$ and $c)$~$\Sigma 3~(211)\langle 011\rangle$ grain boundaries in silicon. The colour of atoms corresponds to segregation energy values given in eV}
\end{figure}

The segregation energy here is the difference between defect formation energies on the GB and in the bulk, i.e. negative value of segregation energy means that segregation is energetically favourable.  As a result, we have built the segregation maps for vacancy on these three boundaries (fig.~\ref{fig:smapV5}). It appears that segregation of vacancy is favourable in most of boundary sites, with lowest segregation energy of -1.8~eV. These results agree with the conclusions of Ref.~\onlinecite{vac-3-112} where the vacancy at $\Sigma 3~(112)$ GB was studied.

\begin{figure}
\center
\includegraphics[width=\columnwidth]{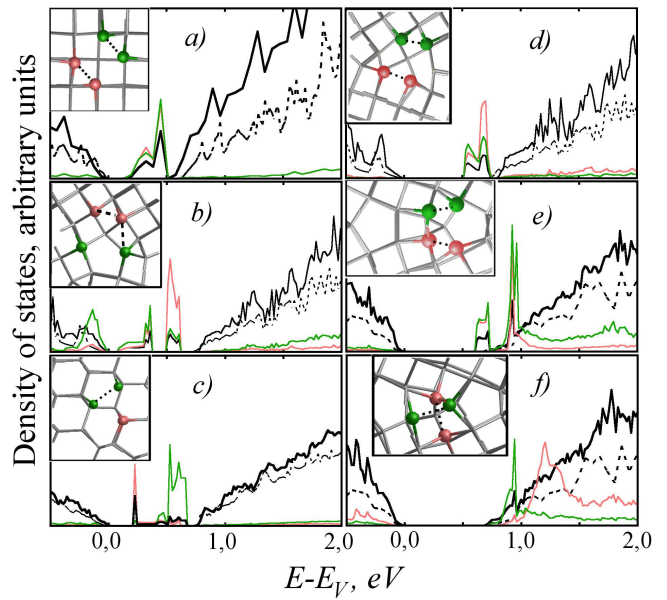} 
\caption{\label{fig:vacancies} Density of states and local geometry (shown in the insets) of $a)$ bulk vacancy; $b)$-$f)$ vacancies on the GB. Dashed line shows the total DOS of the grain boundary without vacancy, thick solid line - total DOS in the presence of vacancy. Thin colour lines present PDOS for the vacancy coordination atoms shown with corresponding colour. Dotted lines on insets show the minimal distances between vacancy coordination atoms.}
\end{figure}

From the variety of possible vacancy-on-the-GB configurations we have selected five which were, on our opinion, the most representative. They cover segregation energy range from -1.8 to 0.2~eV. For selected vacancies we have calculated total density of states (DOS) and projected density of states (PDOS). All results are presented on fig.~\ref{fig:vacancies} together with vacancy geometries shown on insets. The same configurations are characterised in Table~\ref{tbl:vacancies}

\begin{table}
\caption{Calculated parameters of selected vacancy configurations}
\label{tbl:vacancies}
\begin{ruledtabular}\begin{tabular} {c c c c c c c c c}
   & GB & $E_S$, eV & $\eta_1$ & $\eta_2$ & $r_1$,\AA & $r_2$,\AA & $\epsilon$, eV \\
\hline
$a)$ & bulk    & 0.00  & 1.000 & 0.000 & 2.90 & 2.90 & 0.33, 0.44 \\
$b)$ & $(130)$ & 0.20  & 0.987 & 0.003 & 2.95 & 3.12 & 0.35, 0.53 \\
$c)$ & $(211)$ & -0.16 & 1.000 & 0.003 & 2.87 & 3.22 & 0.22, 0.52, 0.60 \\
$d)$ & $(130)$ & -0.69 & 0.978 & 0.002 & 2.67 & 2.86 & 0.53, 0.70 \\
$e)$ & $(520)$ & -1.18 & 0.953 & 0.034 & 2.69 & 2.69 & 0.64, 0.82, 0.96 \\
$f)$ & $(520)$ & -1.80 & 0.918 & 0.016 & 2.50 & 2.58 & 0.93, 1.20 \\
\end{tabular}
\end{ruledtabular}
\end{table}

The first configuration in both fig.~\ref{fig:vacancies} and Table~\ref{tbl:vacancies} is the bulk vacancy. Its segregation energy $E_S$ is zero by definition. Relative tetrahedral volume, i.e. the volume of tetrahedron formed by vacancy coordination atoms {\it before} the segregation divided by the same volume evaluated for perfect bulk is denoted $\eta_1$. For bulk it equals 1, again, by definition. Another local geometry parameter $\eta_2$ will be introduced in section~\ref{sec:params}. $r_1$ and $r_2$ are two nearest distances between vacancy coordination atoms {\it after} geometry optimisation. The approximate positions of vacancy gap states relative to the valence band ceiling ($\epsilon$) are defined as maxima of corresponding peaks of PDOS of vacancy coordination atoms.

There are several clear trends to be observed in Table~\ref{tbl:vacancies}:
\begin{enumerate}
\item There is a correlation between the segregation energy and relative local volume $\eta_1$. In general, the vacancies obey an intuitively clear rule: the more certain atom is squeezed with its neighbours, the more favourable the segregation of vacancy at the position of this atom becomes. 
\item The lower segregation energy leads to the higher positions of vacancy localised levels. Actually, for configurations e) and f) these levels are not even present in the band gap. PDOS of coordination atoms shows that local levels are shifted into conductance band and, hence, become quasilocal.
\item The lower segregation energies and higher local levels position correspond to smaller $r_1$ and $r_2$ distances. These distances are shown with dashed lines of fig.~\ref{fig:vacancies} insets. For the most favourable configuration these distances are 2.50 and 2.58~\AA, i.e. within the covalent bonding range. One can see that as reconstruction of bonds between vacancy neighbours which thus become four-fold coordinated and hence, the structure of boundary with vacancy becomes perfectly bonded. 
\end{enumerate}

In other words, in the favourable positions of vacancy its coordination atoms split into two pairs and atoms of each pair tend to move closer to each other. Those pairs can be seen as having partially restored covalent bonds, and the electronic structure of such bonds becomes similar to that of perfect bulk for smaller interatomic distances. At the same time, total energy gain associated with partial restoration of covalent bonds explains why such configurations are favourable.

The lowest energy vacancy configuration is an example of a problem mentioned in the introduction: in irregular structure it is not always clear what is and what is not ``defect''. Here the perfectly bonded grain boundary segregated the neutral vacancy which is definitely a point defect when considered in the bulk. However, atomic relaxation of vacancy coordination resulted in the structure which is perfectly bonded again and can be seen as yet another possible configuration of the same grain boundary {\it without} coordination defects.

\subsection{\label{sec:impurities} Carbon, phosphorus and boron impurities}

There was certain controversy about boron segregation at grain boundaries in silicon. Segregation of $n$-type impurities (As and P) but no segregation of boron has been observed with laser-assisted 3D atom probe~\cite{dopdistrib-inoue}. Boron impurity segregation, however, has been reported at grain boundary triple points~\cite{thompson-triple}, at $\langle 113 \rangle$ planar defects and dislocation loops~\cite{duguay} where it can even appear as a Cottrel atmospheres~\cite{blavette}. At very high annealing rate (flash annealing) boron has been observed to segregate moderately at grain boundaries~\cite{jin-bsegreg}. At very high concentrations B has been seen to form precipitates at the grain boundaries rather than to segregate there~\cite{pearson}. In the computational study by Arias and Joannopoulos~\cite{arias-joannopoulos} general conclusion has been suggested that $n$-type but not $p$-type impurities segregate at grain boundaries in germanium and silicon.

\begin{figure}
\center
\begin{tabular}{ccc}
\includegraphics[width=0.333\columnwidth]{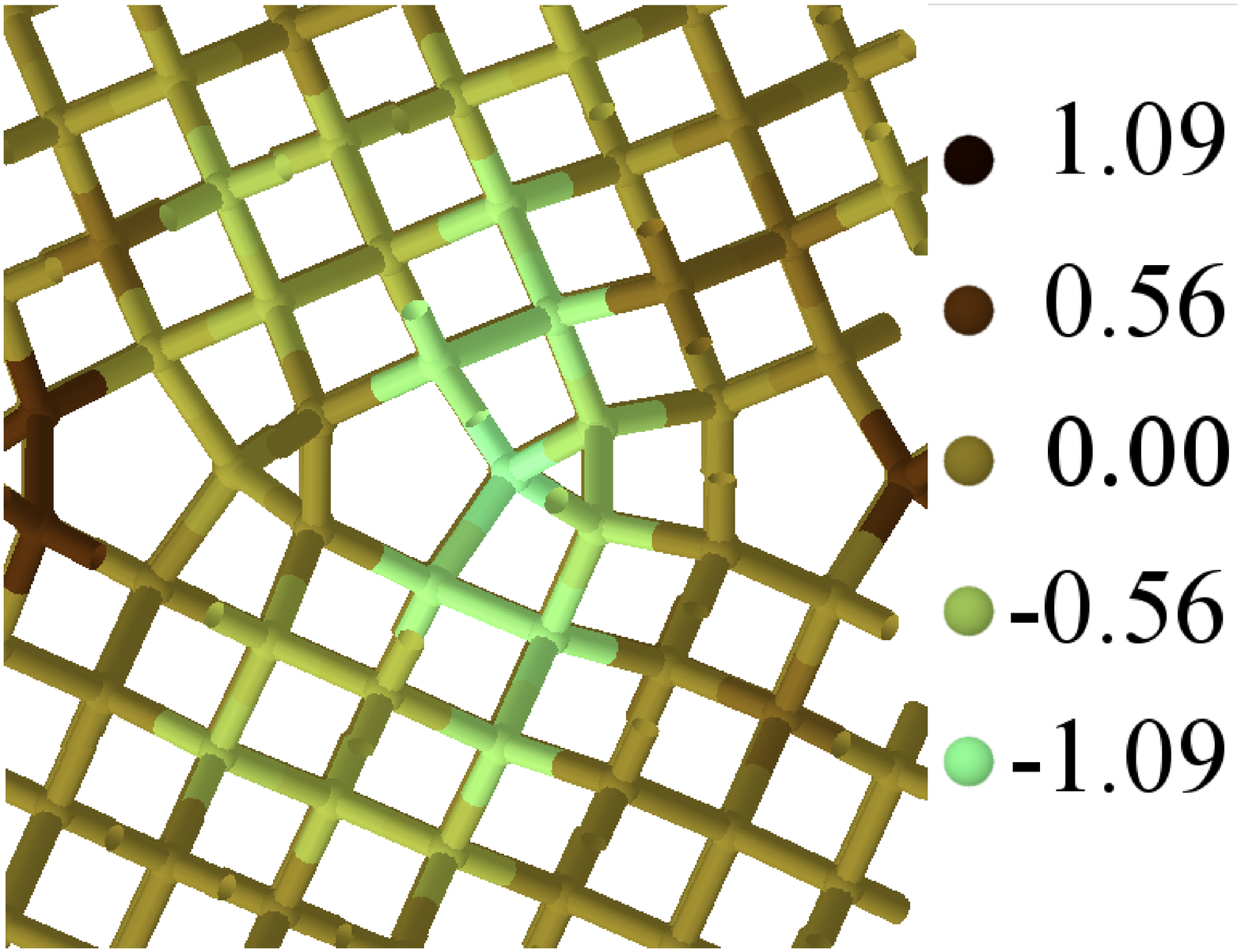} &
\includegraphics[width=0.333\columnwidth]{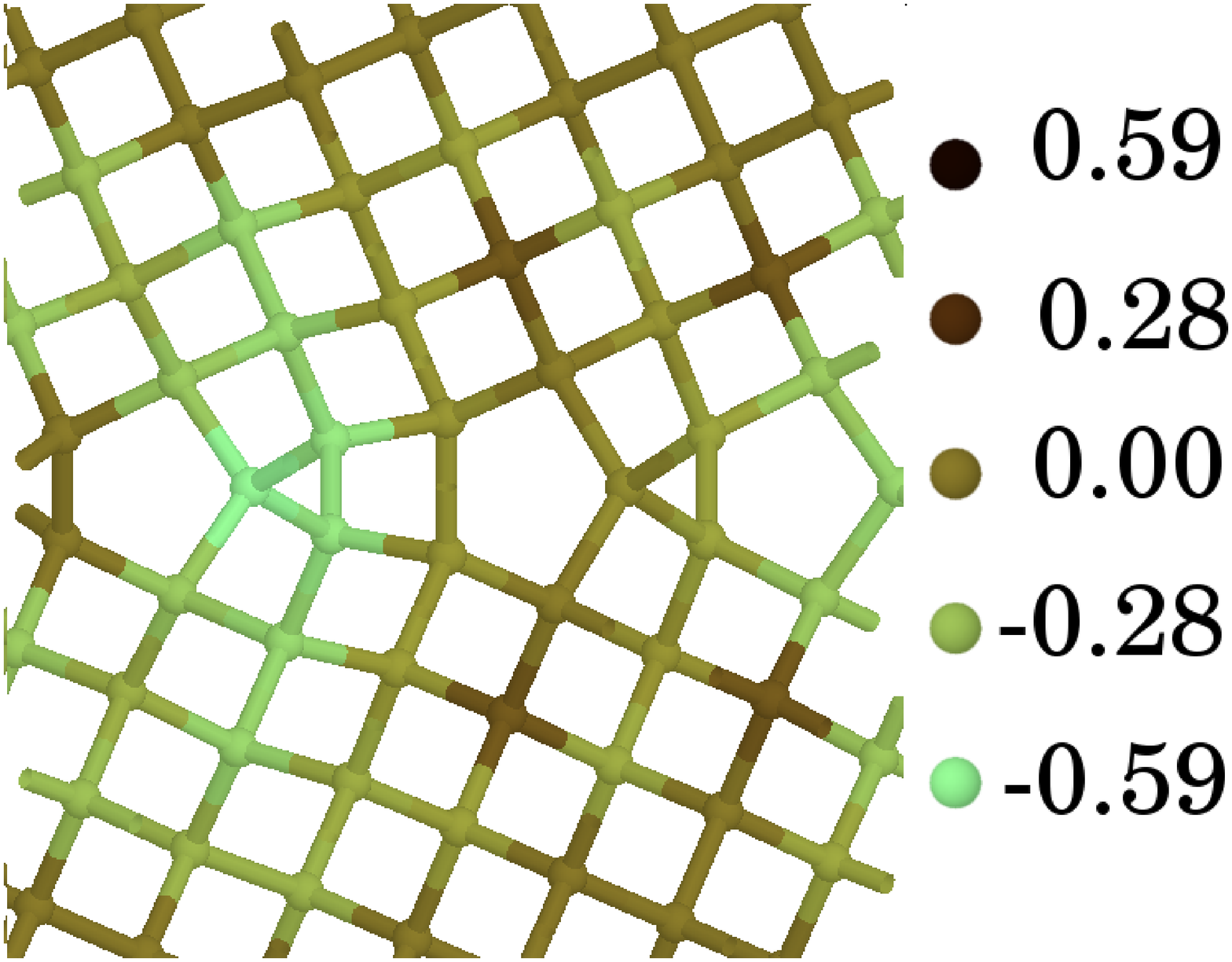} &
\includegraphics[width=0.333\columnwidth]{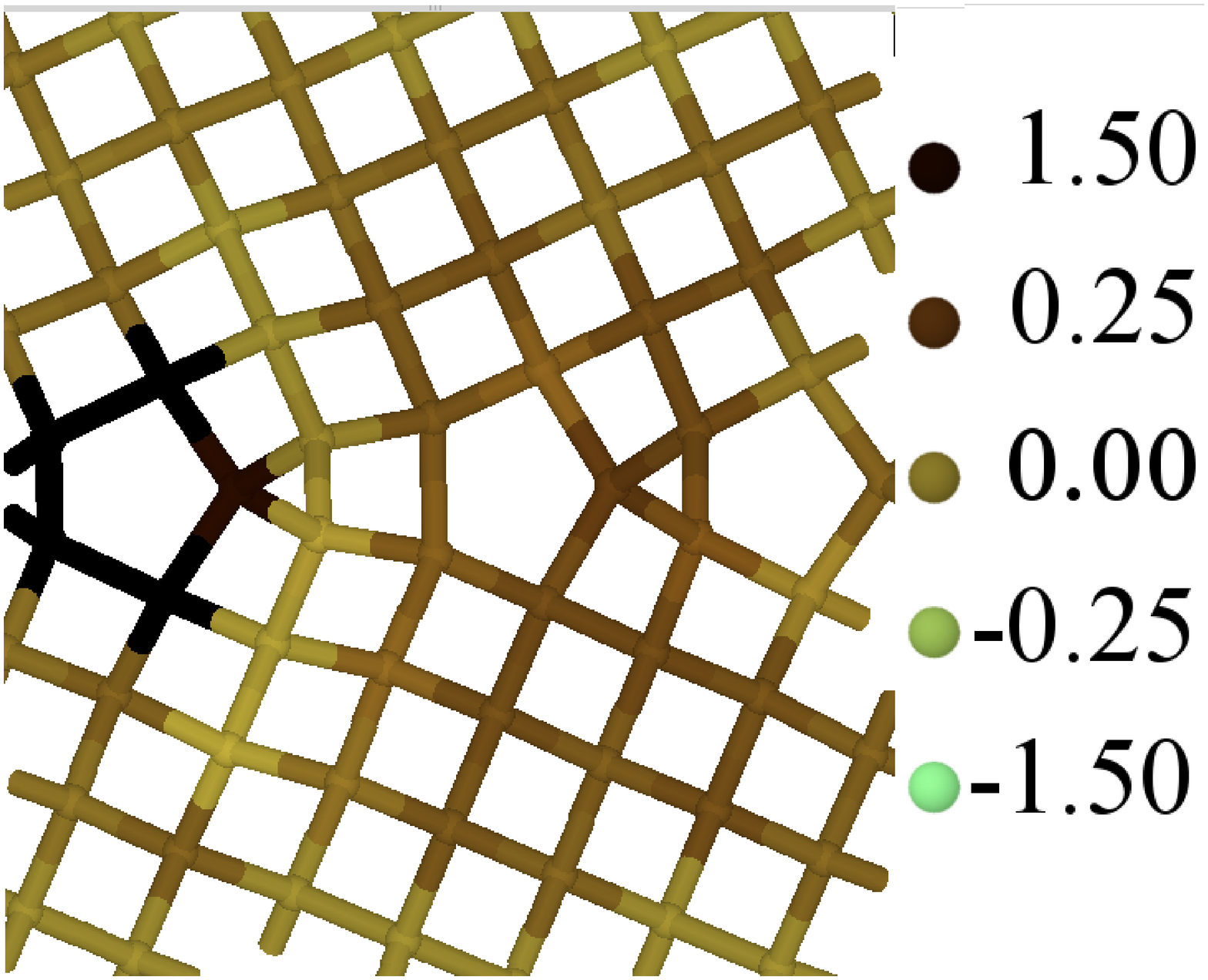} \\
$a)$ & $b)$ & $c)$\\
\end{tabular}
\caption{\label{fig:impmap} Segregation energy maps for $a)$ carbon, $b)$ phosphorus and $c)$ boron impurity at  $\Sigma 29~(520)$ grain boundary }
\end{figure}

For C, P and B impurities we have also calculated the segregation maps, i.e. segregation energies at every site of the $\Sigma 29~(520)$ grain boundary (fig.~\ref{fig:impmap}). Carbon impurity has been treated at atomistic level with Tersoff potentials while phosphorus and boron have been calculated {\it ab initio}. The lowest segregation energies obtained are -1.09, -0.60 and -0.05~eV for C, P and B, respectively. The corresponding highest segregation energies are 0.57, 0.19 and 1.39~eV. The rest of energies for each impurity is distributed approximately uniformly between minimum and maximum values.

One can see from fig.~\ref{fig:impmap} that for carbon and phosphorus there are many favourable segregation sites on the calculated boundaries while for boron there are virtually no such sites. These results qualitatively agree with the suggestion that $n$-type but not $p$-type impurities segregate at grain boundaries in silicon.

\begin{figure}
\center
\includegraphics[width=\columnwidth]{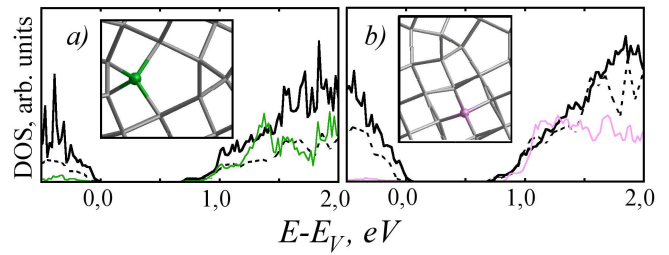} 
\caption{\label{fig:impurity} DOS and PDOS of $a)$ phosphorus and $b)$ boron at $\Sigma 29~(520)$ grain boundaries. Dashed line - DOS of the pure boundary; thick solid line - DOS of impurity-containing boundary; thin colour line - PDOS of impurity atom. Insets show the local geometry of the impurities. }
\end{figure}

 The electronic structure of segregated P and B impurities  brings no surprises. Typical DOS of both impurities is shown on fig.~\ref{fig:impurity}. No deep electron or hole levels are associated with segregated impurities. The question about shallow electronic traps discussed in Ref.~\onlinecite{arias-joannopoulos} requires, on our opinion, more extensive calculations including electronically excited states of the system. Reliable conclusions about the existence of such states can hardly be done on the basis of subtle (about 0.01~eV) energy differences in one-electron spectrum.

\subsection{\label{sec:params}Local deformation parameters}

The working hypothesis behind the following step was simple: we assumed that segregation energy of the defect on certain boundary site depends mainly on the local geometry of that site. Such an assumption should be reasonable, at least, while the defects with delocalised electronic states or long range lattice distortion are not involved. However, even for neutral vacancy which brings long range deformation into the lattice we have already seen the correlation between its properties and relative tetrahedral volume. ``Local'' geometry here means nearest neighbourhood of the defect. As all the defects considered in this paper substitute one silicon atom, it should be sufficient to consider this atom and its tetrahedral coordination.

Let $\bm{r}_i(i=1..4)$ be the positions of four neighbours relative to certain boundary atom. Vectors $\bm{r}_i$ put together make a 12-dimensional vector fully characterising the deformation and orientation of the tetrahedron. The 12-dimensional space of such vectors can be decomposed into subspaces of irreducible representations of $T_d$ symmetry group as $A_1+E+T_1+2T_2$. The $T_1$ component is purely rotational and  irrelevant to the description of the tetrahedron deformation. As for other components, the norm of the deformation vector projection into subspace of each irreducible representation is a good invariant parameter characterising the local deformation.

Finally, the following set of parameters was constructed:
\begin{eqnarray}
\begin{array}{rcl}
\eta_1 &=& \displaystyle\frac{V}{V_0} =  \displaystyle\frac{1}{6V_0}[\bm{r}_{12}\bm{r}_{13}\bm{r}_{14}] \\
a^4\eta_2^2 &=& \displaystyle \frac{1}{4}\sum_i \bm{r}_i^4 - \left( \frac{1}{4}\sum_i \bm{r}_i^2\right)^2 \\
a^4\eta_3^2 &=& \displaystyle (S_{12}-S_{34})^2+(S_{13}-S_{24})^2+\\
 & & +(S_{14}-S_{23})^2\phantom{\sum x} \\
a^4\eta_4^2 &=& \displaystyle (\bm{r}_{12}\bm{r}_{34})^2+(\bm{r}_{13}\bm{r}_{24})^2+(\bm{r}_{14}\bm{r}_{23})^2 \phantom{\sum x}
\end{array}
\label{params}
\end{eqnarray}
Where
\begin{eqnarray*}
\begin{array}{rcl}
\bm{r}_{ij}&=&\displaystyle\bm{r}_j-\bm{r}_i\\
 S_{ij}&=&\displaystyle\bm{r}_i^2+\bm{r}_j^2-6\bm{r}_i\bm{r}_j \phantom{\sum x}\\
a&=&\mbox{ideal bulk Si-Si separation}\\
V_0&=&\displaystyle\frac{8a^3}{9\sqrt{3}}=\mbox{ideal bulk tetrahedron volume}
\end{array}
\end{eqnarray*}
 The tetrahedral volume $\eta_1$ corresponds to $A_1$ irreducible representation of $T_d$ group and, hence, should be  related to the isotropic deformation of tetrahedron. There are many ways to construct a parameter satisfying this demand, for example, mean bond length $\frac{1}{4}\sum_i r_i$ is one of options. However, it has been seen from numerical analysis that tetrahedral volume correlates with segregation energies better than anything else.

Parameters $\eta_2$ and $\eta_3$  both correspond to the $T_2$ irreducible representation and, therefore, their separation is not unique. To make it unambiguous we demanded that $\eta_2$ describes radial deformation (i.e. change in $r_i$ bond lengths) while $\eta_3$ is related to angular distortion of the tetrahedron, i.e. distortion of the bond angles. The latter also applies to the last $\eta_4$ parameter belonging to $E$ irreducible representation.  

\subsection{\label{sec:correl} Correlation between local structure and defect segregation}

\begin{figure}
\center
\includegraphics[width=\columnwidth]{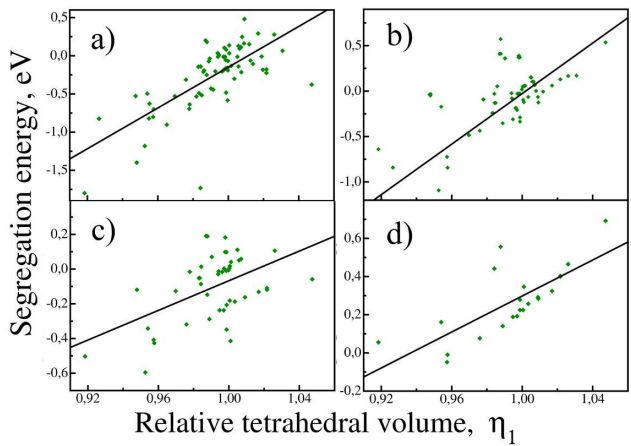} 
\caption{\label{fig:eta1cor} Correlation of the segregation energy of a) vacancy, b) carbon, c) phosphorus and d) boron with relative tetrahedral volume ($\eta_1$ parameter) for  $\Sigma 5~(130)$, $\Sigma 29~(520)$ and $\Sigma 3~(211)$ GBs. Solid line shows the least square linear fit.}
\end{figure}

We have already observed in section~\ref{sec:vacancy} the correlation between segregation energy of neutral vacancy and relative tetrahedral volume $\eta_1$. The same correlation exists for C, P and B impurities as well (fig.~\ref{fig:eta1cor}). However, correlation with $\eta_1$ is the strongest but not the only one.

\begin{figure}
\center
\includegraphics[width=\columnwidth]{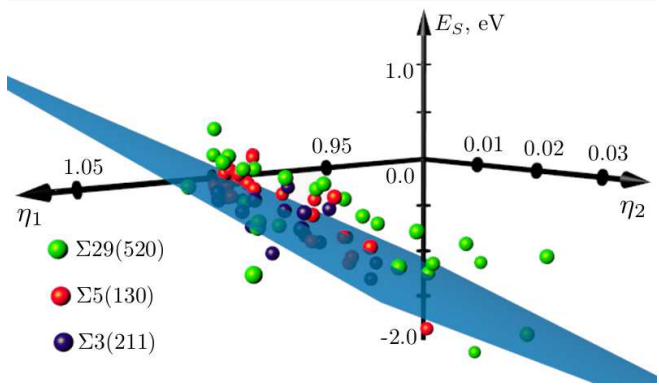} 
\caption{\label{fig:3dcor} 3d-view of the correlation between neutral vacancy segregation energy and $\eta_1$, $\eta_2$ parameters for $\Sigma 29 (520)$, $\Sigma 5 (130)$ and $\Sigma 3 (211)$  GBs}
\end{figure}

We have analysed the correlation between segregation energies of calculated defects and $\eta_{1-4}$ parameters. The important point is that the parameters are calculated {\it before} the defect is introduced in some lattice site.  The search for correlation have been performed in simplest linear form:
\begin{equation}
 E_S = E_{S0}+a(\eta_1-1)+b\eta_2+c\eta_3+d\eta_4
\label{segreg-correl}
\end{equation}
The  coefficients $E_{S0}$, $a$, $b$, $c$, $d$ obtained by least square fit are shown in Table~\ref{tbl:coeffs}. It appears that besides relative tetrahedral volume $\eta_1$ segregation energies of defects do correlate with other local deformation parameters. For example, vacancy segregation energy exhibits the correlation with $\eta_2$ but no correlation with $\eta_3$ and $\eta_4$. Here the judgement about presence or absence of correlation is being done on the basis of relative fitting error for the corresponding coefficients - if it exceeded 30\% we  considered that as little or no correlation. Fig.~\ref{fig:3dcor} gives the example of the three-dimensional correlation view for the vacancy. One can see that the points representing computed segregation energies are situated near the plane built according to the formula~\ref{segreg-correl}. For C, P and B impurities the same type of picture can be built an it looks very similar - segregation energies group near the plane with only few distant points.

\begin{table}
\caption{The coefficients of approximate  segregation energy function. The relative fitting error is given in parentheses}
\label{tbl:coeffs}
\begin{ruledtabular}\begin{tabular} {c c c c c c c c c}
 Defect & $E_{S0}$, eV & $a$, eV & $b$, eV & $c$, eV & $d$, eV \\
\hline
Vacancy &  $\begin{array}{c} 0.07 \\ (75\%)\end{array}$  & $\begin{array}{c} 11.9 \\ (12\%) \end{array}$   & $\begin{array}{c} -17.6 \\ (20\%) \end{array}$  & --   & -- \\

C       & -0.03    & 13.9     & -12.6    & --   & 0.9 \\
        & (105\%)  & (6\%)    & (15\%)   &      & (8\%) \\
P       & -0.02    & 4.2      & -11.0    & 0.13 & -- \\
        & (162\%)  & (19\%)   & (15\%)   &(22\%)&  \\

B       & $\begin{array}{c} 0.19 \\ (8\%)\end{array}$    
&  $\begin{array}{c}  6.5 \\ (6\%) \end{array}$   &   --     & --   
& $\begin{array}{c} 0.4 \\ (10\%)\end{array}$ \\
\end{tabular}
\end{ruledtabular}
\end{table}

The presence of non-zero $E_{S0}$ term in the approximation (\ref{segreg-correl}) might be considered discouraging. Indeed, the values $\eta_1=1, \eta_2=\eta_3=\eta_4=0$ correspond to perfect bulk geometry and segregation energy of any defect in such situation should be zero by definition. However, the values of $E_{S0}$ obtained for the vacancy as well as for C and P impurities are small and have large fitting error. This means those values are actually undefined but small in perfect agreement with the demand $E_{S0}=0$. The only exception is B impurity, for which we have to admit the presence of systematic error in calculated segregation energy. 



\section{\label{sec:concl}Conclusion}

In present paper we have reported the results of atomistic and {\it ab initio} simulation of several symmetric and asymmetric tilt boundaries in silicon. We have also investigated the connection between defect segregation on the boundary and its local structure.

The implementation of genetic algorithm we have applied to GB structure search was able to reproduce structures of $\Sigma 5 (130)$, $\Sigma 3 (211)$ and $\Sigma 29 (520)$ boundaries. However, some of attempted GA runs were unsuccessful and one even ended up with amorphous-like energetically unfavourable configuration of $\Sigma 29 (520)$ GB, although perfectly bonded. This demonstrates that even a sophisticated tool like GA still is not fully reliable and universal for searching GBs lowest energy structures.

Perfectly bonded structures have been found for asymmetric  $\Sigma 9  (\overline{2}55)/(\overline{2}11)\langle 011\rangle$, $\Sigma 3(\overline{2}55)/(211)\langle 011\rangle$, $\Sigma 13  (790)/(3~11~0) \langle 001\rangle$ GBs. The zigzag structure of $\Sigma 3(\overline{2}55)/(211)\langle 011\rangle$ GB appeared to have surprisingly low energy of 0.37~$J/m^2$ which makes it a good candidate for possible experimental observation. For the other two GBs the energies were about 1~$J/m^2$. Taking into account the above comment on GA reliability, it cannot be stated for sure these structures are lowest energy ones. Even if they are not, however, our results strongly support the position that pure tilt boundaries in silicon (both symmetric and asymmetric) do not have coordination defects and associated gap states.

The main result of the present paper is the unambiguous correlation between segregation energies of both intrinsic and impurity defects and local geometry of the boundaries. We have introduced purely geometric parameters characterising the local structure of a boundary site and determined correlation coefficients between segregation energies and these parameters. It is quite possible that similar correlations might be found not only in tilt boundaries, but also in other boundary types, in dislocations and even in amorphous silicon. The approximate formula~\ref{segreg-correl} can be used for semiquantitative evaluation of segregation energies and requires only the knowledge of GB structure. This might potentially greatly facilitate:

\begin{itemize}
\item Determination of boundary types which can actively segregate certain defect;
\item Determination of favourable boundary sites for the segregation of certain defect;
\item Modelling the equilibrium distribution of  impurities between the GB and the bulk;
\item Kinetic Monte-Carlo simulations of the diffusion and the segregation of impurities in polycrystalline silicon.
\end{itemize}

The present work has revealed several features of the neutral vacancies on the boundary. With the segregation energy decreasing from 0 to -1.8~eV the deep levels of the vacancy shift towards the conduction band, become shallow levels and then, for the most energetically favourable positions of vacancy, disappear inside the conduction band. At the same time the atoms of vacancy coordination form pairs with decreasing interatomic distances. For the most favourable positions one can state that covalent bonds between these atoms are reconstructed. In other words, the migration of vacancy into the most favourable boundary sites can lead to the structural transformation of the boundary, while the vacancy itself ``disappears'', leaving perfectly bonded structure with no local states in the band gap.

\section*{\label{sec:akn}Acknowledgement}
 
The authors gratefully appreciate the use of computational clusters located in Institute for System Dynamics and Control Theory SB RAS and in Novosibirsk State University Supercomputer Centre. The study was supported by The Ministry of education and science of Russian Federation.


\bibliography{article}

\end{document}